\documentclass[preprint2]{aastex}

\lefthead{Lipari et al.}

\righthead{GALACTIC-WINDS, MERGERS AND IR QSOs}

\begin{document}

\title{EXTREME GALACTIC-WINDS AND STARBURST \\
 IN IR MERGERS AND IR QSOs\altaffilmark{1}}

\vspace{3mm}

\altaffiltext{1}{Based on observations obtained at the 
              Bosque Alegre Station  (BALEGRE) Argentina,
              Complejo Astronomico El Leoncito (CASLEO) Argentina,
              European Southern Observatory (ESO) Chile,
              Hubble Space Telescope (HST-WFPC2) satellite, 
              Kitt Peak National Observatory (KPNO) USA,
              Mauna Kea Observatory, Hawaii (MKO) USA,
              Palomar Observatory USA.}

\author{S. L\'{\i}pari\altaffilmark{2,3},
        D. Sanders\altaffilmark{3,4},
        R. Terlevich\altaffilmark{5}, 
        S. Veilleux\altaffilmark{3,6},
        R. D\'{\i}az\altaffilmark{2,3},  
        Y. Taniguchi\altaffilmark{7}, \\
        W. Zheng\altaffilmark{3,8},
        D. Kim\altaffilmark{3,9},  
        Z. Tsvetanov\altaffilmark{8},
        G. Carranza\altaffilmark{2}, and
        H. Dottori\altaffilmark{10}}

\altaffiltext{2}{Cordoba Observatory and CONICET, Laprida 854, 5000 Cordoba,
Argentina (lipari@mail.oac.uncor.edu).} 

\altaffiltext{3}{Visiting astronomer at BALEGRE, CASLEO, ESO, KPNO, MKO
Observatories.}

\altaffiltext{4}{Institute for Astronomy, University of Hawaii, 2680 Woodlawn
                 Drive Honolulu, HI 96822 USA.}

\altaffiltext{5}{Institute of Astronomy, Madingley Road, Cambridge CB3 0HA,
                  United Kingdom.}

\altaffiltext{6}{Department of Astronomy, University of Maryland, College Park,
                 MD 20742 USA.}

\altaffiltext{7}{Astronomical Institute, Tohoku University, Aoba, Sendai
                 980-8578, Japan.}

\altaffiltext{8}{Department of Physic \& Astronomy, University of Johns Hopkins, 
                 Baltimore, MD 21218, USA.}

\altaffiltext{9}{Academia Sinica, Institute of Astronomy \& Astrophysics,
                 PO Box 1-87, NanKang, Taipei 115, Taiwan.}

\altaffiltext{10}{Instituto de Fisica, Univ. Fed. Rio Grande do Soul, CP
                 15051, Porto Alegre, Brazil.}

\begin{abstract}

We report --as a part of a long-term study of {\it mergers and IR QSOs}--
detailed spectroscopic evidences for outflow (OF) and/or Wolf Rayet (WR)
features in: (i) low velocity OF ongoing mergers 
NGC 4038/39 and IRAS 23128-5919; and
(ii) extreme velocity OF (EVOF) QSOs IRAS 01003-2238 
and IRAS 13218+0552. We also study the presence of OF and EVOF in a complete sample
of ultra-luminous IR galaxies/QSOs (``The IRAS 1 Jy MKO-KPNO Survey", of 118
objects).
We found EVOF in IRAS 11119+3257, 14394+5332, 15130+1958 and 15462-0450
(and probable OF  in IRAS 05024-1941, 13305-1739, 13451+1232,  and
23389+0300).
The OF components detected in these objects were mainly associated to starburst processes: i.e.,
to galactic--winds generated in multiple type II SN explosions and massive stars.
The EVOF were detected in objects with strong starburst plus obscured IR QSOs;
which suggest that interaction of both processes could generate EVOF.

In addition, we analyze  the presence of Wolf Rayet  features
in the large sample of Bright PG-QSOs (Boroson and Green 1992), and
nearby mergers and galactic-wind galaxies.
We found clear WR features in the Fe II PG-QSOs (type I):
PG 1244+026, 1444+407, 1448+273, 1535+547; and in the IR merger Arp 220.
We describe the properties of the [O III]$\lambda$5007-4959 emission,
in strong and extreme  Fe\,II+IR+BAL emitters (QSOs of types I and II).

HST archive images of IR+BAL QSOs show in practically all of these objects
{\it ``arc or shell"} features
probably associated to galactic-winds (i.e., to multiple type II SN explosions)
and/or merger processes.

Finally,
we discuss the presence of extreme starburst and galactic wind as a possible
evolutive link between IR merger and IR QSOs; where
the relation between {\it mergers and extreme starburst (with
powerful galactic--winds and ``multiple" type II SN explosions)} plays an
important role, in the evolution of galaxies/QSOs.

\end{abstract}

\keywords{ 
galaxies: individual (IRAS 00275-2859, IRAS 01003-2238, IRAS 04505-2958,
IRAS 07598+6508, IRAS 11119+3257, IRAS 13218+0552, IRAS 14026+4341,
IRAS 14394+5332, IRAS 15130+1958, IRAS 15462-0450, IRAS 19254-7245,
IRAS 23128-5919, Mrk 231, NGC~4039/38, PG 1244+026, PG 1448+273,
PG 1444+407, PG 1535+547, PG 1700+518, and I Zw 1)
-- galaxies: interactions -- galaxies: starburst -- quasar: general  
}

\section{INTRODUCTION}

Main current issues in astrophysics are the study of
{\it mergers, massive star formation processes, IR/BAL
QSOs (and the relation between them)}.
These three issues play an important
role in practically all the scenario of {\it formation and evolution  of
galaxies and active galactic nuclei (AGNs)} (see for references \S 4.4).
In addition, the presence of galactic-winds --associated mainly to extreme star
formation processes-- is also a main component for different theoretical
models  of {\it galaxies formation} (see Ostriker \& Cowie 1981; Berman \&
Suchkov 1991).

{\it Mergers} are mainly luminous IR galaxies whose luminosities overlap with 
most luminous QSOs and Seyfert galaxies; and their optical, IR and radio properties
show starburst and AGN characteristics (Joseph \& Wright 1985;
Schweizer 1980, 1982, 1996; Rieke et al. 1985; Sanders et al. 1988a;
Sanders \& Mirabel 1996).
{\it Luminous IR galaxies} (L$_{IR} \geq 10^{11} L_{\odot}$; LIRGs)
are dusty strong IR emitters where an enhancement of {\it star formation}
is taking place (see for references Lipari et al. 2000a); and
imaging surveys of ultraluminous IR galaxies
(L$_{IR} \geq 10^{12} L_{\odot}$; ULIRGs) show that $\sim$100$\%$ 
are mergers or strongly interacting systems 
(Sanders et al. 1988a; Melnick \& Mirabel 1990; Clements et al. 1996).
And, there is compelling evidence for
merger and interaction driven {\it starburst and nuclear activity}; probably by 
depositing large amounts of interstellar gas to the nuclear regions (see 
Sanders \& Mirabel 1996; Scoville \& Soifer 1991; Barnes \& Hernquist 1992; 
Mihos \& Hernquist 1996, 1994a,b). In addition {\it galactic--winds,
bubbles and Wolf Rayet} features have been clearly detected, in starburst
and IR galaxies (see for references Lipari et al. 2000a; Heckman, Armus, \&
Miley 1990, hereafter HAM90; Lipari \& Macchetto 1992).

The discovery and study of luminous {\it IR QSOs and IR selected QSOs}
(see Beichman et al. 1986; Vader et al. 1987; Lawrence et al. 1988; Sanders
et al. 1988b; Low et al. 1988, 1989; Lipari, Macchetto, \& Golombeck 1991;
Colina, Lipari \& Macchetto 1991b; Lipari, Colina \& Macchetto 1994; and others)
raises several interesting questions, in particular: are they a new or a
special class of QSOs?.
In the last years, it has been proposed that luminous IR QSOs are normal AGNs,
merely viewed at particular angles (Wills et al. 1992; Hines \& Wills 1995;
Boyce et al. 1997).
While this hypothesis is interesting, it has problems to explain several 
observational results  (Canalizo, Stockton \& Roth 1998). 
On the other hand,
we found that almost {\it 100\% of extremely strong Fe II emitters} are luminous IR 
{\it QSOs} and are  radio-quiet. We suggested that these objects would be young IR active galaxies at the end phase of a strong starburst
(Lipari, Terlevich \& Macchetto 1993; Lipari 1994).
In recent years,  {\it composite models} (i.e., nonthermal--AGN plus
starburst/superwind scenario) have become a widely accepted approach
in studying the source of nuclear energy in IR galaxies and QSOs (see Genzel et al.
1999; Downes et al. 1999; Veilleux, Kim \& Sanders 1999; Lutz, Veilleux,
 \& Genzel 1999; Schmit el al. 1998).

In addition, some of the results 
obtained for low ionization BAL QSOs, such as very weak [OIII]$\lambda$5007 
emission, strong blue asymmetry in H$\alpha$, radio quietness, and strong 
IR and Fe II emission (Boroson \& Mayer 1992; Lipari et al. 1993, 1994),
 can be explained in the framework of the starburst scenario (Lipari 1994).
In our study of Mrk 231, IRAS 0759+6559 (the nearests extreme Fe II, IR and
BAL systems), we detected typical characteristics of a young-starburst QSO
and found evidence of a probable link between BAL systems and star formation
 regions (Lipari et al. 1994; Lipari 1994; Dyson et al. 1992).
Specifically, our evolutive model for young IR QSOs (see for references
Lipari 1994) suggested that BAL systems
could be linked to violent supermassive starburst which produce an expanding
shell and dust obscuration. Recently, several articles suggested that this
evolutive model shows a good agreement  with the observations (see
Canalizzo, Stockton \& Roth 1998, for references).

In the last years,
several possible links between {\it mergers, starburst/galactic-wind,
and IR~QSOs/galaxies} have been proposed.
Specifically, Joseph et al., Sanders et al. and Lipari et al. suggested
three complementary sequences and evolutive--links:

(i) merger $\to$ giant shocks $\to$ super-starbursts + galactic winds 
$\to$ elliptical galaxies;

(ii) merger $\to$ H$_2$-inflow (starbursts) $\to$ cold ULIRGs $\to$
  warm ULIRGs + QSOs;

(iii) merger{\bf /s} $\to$  extreme starburst + galactic-wind
(inflow + outflow) $\to$ extreme

Fe II/BAL/IR composite-QSOs   $\to$ standard QSOs and elliptical.

Studies  of nearby mergers and IR QSO with high resolution is one
observational way to study this relation.
We have started a detailed {\it morphological, spectroscopic and
kinematical} study of luminous mergers and IR QSOs (Lipari et al. 2000a); 
and we present -in this paper- several selected results from this program.
We analyze in this paper the OF and WR features in:
(i) two nearby mergers, the ``Antennae"
(z = 0.0055, L$_{IR[8-1000 \mu m]} \sim 1.0 \times 10^{11}
L_{\odot}$); and IRAS 23128-5919
(z = 0.044, L$_{IR[8-1000 \mu m]} \sim 2.0 \times 10^{12} L_{\odot}$);
and (ii) two relatively nearby IR QSOs,
IRAS 01003-2238 (z = 0.118, L$_{IR[8-1000 \mu m]} =
1.74 \times 10^{12} L_{\odot}$) and
IRAS 13218+0552 (z = 0.20, L$_{IR[8-1000 \mu m]} = 4.26 \times 10^{12}
L_{\odot}$, L$_{IR}$/L$_{B} \sim$ 90; see Armus, Heckman, \& Miley
1988; Sanders et al. 1988b; Low et al. 1988, 1989; Remillard et al. 1993).
In addition, we study EVOF and WR features in the large sample of ultra-luminous
IR galaxies/QSOs (``The IRAS 1 Jy  Survey"; Kim \& Sanders 1998;
Kim, Veilleux \& Sanders 1998; Veilleux, Kim \& Sanders 1999)
and Bright PG-QSOs (of Boroson \& Green 1992), respectively.
Throughout the paper, a Hubble constant of H$_{0}$ = 75 km~s$^{-1}$
Mpc$^{-1}$ will be assumed.

\section{OBSERVATIONS AND REDUCTIONS}

The observations were obtained at BALEGRE, Casleo, ESO, KPNO, MKO, and Palomar
Observatories, with the 1.54 m, 2.15 m, 2.2 m, 3.6 m and 5.0 m telescopes.
In addition, HST--(WFPC2/NICMOS) and ESO archive images were studied. 

Long-slit spectroscopic observations were 
obtained at the 1.54 m telescope of Bosque Alegre Station of Cordoba Observatory
using the Multifunctional Integral Field Spectrograph (Afanasiev, Dodonov \& 
Carranza 1994; Diaz et al. 1999)
during 10 selected photometric nights between 1998 January and 2000 May. 
The observations were made mainly using a slit width of 1.0$''$ 
and 1200 l/mm grating, 
which gives an effective resolution of $\sim$90 km~s$^{-1}$ and a dispersion of 
40 \AA\ mm$^{-1}$ covering the wavelength range 
$\lambda\lambda$6400--6900 \AA.
In order to have accurate spatial positions for the velocity
determinations, zero order imaging of object
plus slit were used.
The seeing was in the range 1.3--2.3$''$  (FWHM).

Long-slit observations were obtained at the 2.15 m telescope of Casleo (San
Juan, Argentina) during the period 1997 June to 2000 May. The observations
were made using the intrumental configuration and technique described in
previous papers (see \S 1, for references).

The data of the IRAS 1 Jy ULIRGs sample were obtained with Gold Cam
spectrograph on the
KPNO 2.15 m telescope using a grating of 300 lines mm$^{-1}$ (8 \AA, of
resolution). And with the Faint Object Spectrograph at the f/10 Cassegrain
focus of the University of Hawaii 2.2 m telescope at Mauna Kea; 
 using a grating of 600 lines mm$^{-1}$ (8 \AA, of resolution).
The details of the spectra obtained at Palomar 5.0 m telescope  were
published by Sanders et al. (1988a).

The ESO Faint Object Spectrograph and Camera (EFOSC) on the 3.6 m telescope at La Silla
was used to obtain long-slit spectra and high resolution images.
Medium-resolution spectra were obtained with the B150, O150, and R150 grisms,
which provide a dispersion of 130 \AA mm$^{-1}$ in a range $\lambda\lambda$
3600-9800.

The HST--WFPC2 observations include broad-band images, using the
filters F450W ($\sim$B Cousins filter) and F814W ($\sim$I); with
a CCD scale  of 0.046$''$ pixel$^{-1}$, in the PC. 
HST--NICMOS observations include near-IR images, using the NIC2 and NIC1
CCD-camera.

The IRAF\altaffilmark{11}, SAO\altaffilmark{12} and ADHOC\altaffilmark{13}
software packages were used to reduce the spectrophotometric data. 
Bias and dark subtraction and 
flat fielding were performed in the usual way. Wavelength calibration of the 
spectra was done by fitting two dimensional polynomials to the position 
of lines in the arc frame. The spectra were corrected for atmospheric 
extinction, galactic reddening, and redshift. The spectra and images 
were flux calibrated using observations of standard stars from the samples 
of Stone \& Baldwin (1983) and Landolt (1992). 
The emission lines were measured and decomposed using Gaussian profiles 
by means of a nonlinear least-squares algorithm described in Bevington (1969).

\altaffiltext{11}{IRAF is the imaging analysis software facility
developed by NOAO}

\altaffiltext{12}{SAO is the imaging analysis software developed by the 
Special Astrophysical Observatory, of USSR Academy of Science}

\altaffiltext{13}{ADHOC is the imaging analysis software developed by 
Marseille Observatory}

We have digitized the published spectra of IRAS 13218+0552  (from
Remillard et al. 1993) and PG 1244+026, 1444+407, 1448+273 and 1535+547 (from
Boroson \& Green 1992).

\section{RESULTS}

In this section we study spectroscopic and morphological
evidence of outflow --associated mainly to galactic-superwind-- in
mergers, IR QSOs, and ULIRGs (where  previously  were detected
starburst and Wolf-Rayet features, bipolar extended  emission,
extreme IR emission, BAL systems, etc.). In addition, we analyze:
(i) the presence of Wolf Rayet features in the sample of Brigth PG QSOs
(Boroson \& Green 1992) and in mergers or galactic-wind galaxies; and
(ii) the host galaxies of IR+BAL+Fe II QSOs (using high resolution HST
images).

It is important to note that the results presented in this paper
are part of a log-term multiwavelength study (including morphological, kinematical
and physical conditions data) of luminous  IR mergers and  IR QSOs
(see Lipari et al. 2000a). These observations were performed mainly in order
to study their links.

\vspace{5mm}
\subsection{{\it Nuclear Outflow and WR features in ``The Antennae"
 and IRAS 23128-5919}}

Read, Ponman \& Wolstencroft (1995) and Fabbiano, Schweizer, \& Mackie (1997)
already suggested the presence of outflow in The Antennae, from x-ray observations.
In order to detect this  component, in the nuclear and central regions
of NGC 4038/39, long exposure spectroscopic observations were performed
at CASLEO, with added long-slit spectra of $\sim$3 hs.  This
spectra were taken mainly through both nuclei. 
We found  in the southern nucleus (NGC 4039) a defined
blue component; which is evident mostly in the H$\alpha$ and
[N II]$\lambda$6584 emission lines (Fig. 1a).
The velocity of this nuclear outflow is V$_{Nucl. OF}$ =
(-365 $\pm$ 50) km~s$^{-1}$.
In \S 4 this component  will be associated  to the nuclear starburst,
detected previously in the nucleus of NGC 4039.
We note that this blue outflow component (with relatively low 
velocity) was clearly detected only using data of  moderate spectral
resolution ($\sim$90 km~s$^{-1}$ FWHM) and high S/N. We found a similar
result in the main optical nucleus of the IR merger NGC 3256
(Lipari et al. 2000a).

We observed for the outflow components similar intensities in 
these emission lines; however, the outfow is stronger in the H$\alpha$ line
than in [N {\sc ii}] emission (we found in NGC 3256 and NGC 4945 the opposite
 result).
It is interesting to note that Casleo spectra of NGC 4039 nucleus
show LINER properties (Lipari et al. 2000b), and  NGC 4038 nucleus show H II
regions features.
The values of the [N {\sc ii}]$\lambda$6584/H$\alpha$,
[O {\sc i}]$\lambda$6300/H$\alpha$ and [S {\sc ii}]$\lambda$6517-31/H$\alpha$
emission line ratios in the nuclear region are clearly consistent with
shocks driven into clouds accelerated outward by a starburst with
galactic--wind (see Lipari et al. 2000b and  HAM90: their Fig. 14).
In addition, is important to note that Rosa \& D'Oddorico (1986) reported the
presence of WR features in the HII regions of the Antennae.

The presence of outflow and WR-features in IRAS 23128-5919 has been
previously suggested (Bergvall \& Johansson 1985; Johansson \& Bergvall 1988).
We found, using CASLEO spectra, low velocity outflow and WR features in the bright
southern nucleus of IRAS 23128-5919, with a value of
V$_{Nucl. OF}$ = (-300 $\pm$ 70) km~s$^{-1}$.

Using spectra of Arp 220 of high S/N (obtained at Palomar 5.0 m
telescope; Sanders et al. 1988) we detected in the bright north-west nucleus
weak (but clear, at S/N $\sim$ 5) WR features at NII$\lambda$4640 and He
II$\lambda$4686.

\vspace{5mm}
\subsection{{\it Outflow and WR features in IR QSOs IRAS 01003-2238
and IRAS 13218+0552}}

IRAS 01003-2238 and IRAS 13218+0552 are probably two of the more interesting
IR QSOs (see \S 1, for references).
In the spectrum of  IRAS 01003-2238 and IRAS 13218+0552
(obtained at MKO  and from Remillard et al 1993; respectively)
we detected  {\it extreme multiple outflow components}, mainly in the
[OIII]$\lambda\lambda$5007-4959 and H$\beta$ emission lines (Figs. 1b and 1c).
We measured for IRAS 01003-2238 outflow-velocities of V$_{OF 1}$ =
(-1530 $\pm$ 60) km s$^{-1}$ and V$_{OF 2}$ = (-710 $\pm$ 50) km
s$^{-1}$, and for the main emission line component (MELC), a value of
cz$_{MELC}$ = 35505 km s$^{-1}$. 
For IRAS 13218+0552 we have obtained  V$_{OF}$ =
(-1850 $\pm$ 90) km s$^{-1}$ 
s$^{-1}$, and cz$_{MELC}$ = 61000 km s$^{-1}$.

In \S 4 these OF components  will be associated  to the nuclear starburst
and the interaction between the starburst+QSO processes.
Again this blue outflow  was clearly detected only using high quality
data.

On the other hand, both IRAS 01003-2238 and IRAS 13218+0552 show
 merger morphology (Surace et al. 1997; Boyce et al. 1997).
Armus et al. (1988) found WR-features in IRAS 01003-2238; and
we detected similar WR feature in IRAS 13218+0552 (using the MKO data, and
our digitized spectra
of the observation published by Remillard et al. 1993).

Finally,
it is important to note that previously we found a very similar result
(i.e., EVOF  in [O III] emission line)  for two
ULIRGs. First, in the southern nucleus
of the IR merger  19254-7245 (the ``super-antennae", with
V$_{syst.}$ = 17900 km s$^{-1}$, L$_{IR}$ = 1.1 $\times$ 10$^{12}$ L$_{\odot}$,
M$_B$ = -23.3 and d$_{nuclei}$ = 10 kpc),  we detected
 a massive star formation process with a strong galactic-wind and
 V$_{OutFlow}$ $\sim$-1000 km s$^{-1}$, plus a type 2 QSO
 (Colina, Lipari \& Macchetto 1991a). We also detected
similar EVOF in Mrk 231 (V$_{OutFlow}$ $\sim$-1000 km s$^{-1}$,
in [O II]), this IR merger also has an obscured QSO (type 1) plus a
strong circumnuclear starburst. In \S 4.2 we comment the properties of these
EVOF objects.

\vspace{5mm}
\subsection{{\it Extreme Outflow in the 1 Jy Sample of ULIRGs}}

An interesting test in order to study {\it the role of starburst/galactic-wind
(GW) in IR mergers and IR QSOs (and their relation)}, is the analysis
of outflow in a complete sample of luminous IR Galaxies and QSOs. Although the
detection of low velocity outflow (V$\leq$ 600 km s$^{-1}$) required
detailed observation (see \S 3.2 and Lipari et al. 2000a).
The study of outflow with large velocities (EVOF) -observed as
strong multiple components in the emission line [O III]$\lambda$5007- requires
mainly spectra with moderate spectral resolution (objects showing
very high FWHM([O III]) values).

Therefore, we performed a detailed study of EVOF, with FWHM ([O III]) $\geq$
1000 km s$^{-1}$, in the MKO-KPNO Survey of 1 Jy ULIRGs and IR QSOs
(Kim \& Sanders 1996; Kim, Veilleux \& Sanders 1998; Veilleux, Kim \& Sanders
1998).
We study the ULIRGs: IRAS 05024-1941,
11119+3257, 132188+0552, 13305-1739, 13451+1232, 14394+5332, 15130-1958,
15462-0450, and 21219-1757. In addition, we study
IRAS 23389+0300 that shows narrow plus broad [O III] components.
And we found interesting results:

\begin{enumerate}
\item

we detected clear EVOF in [O III] and H$\beta$ (see Table 1),  in:

IRAS 11119+3257 (see Fig. 2), IRAS 14394+5332, IRAS  15130+1958,
and IRAS 15462-0450.

\item
and probable OF
in: IRAS 05024-1941, 132188+0552, 13305-1739, 13451+1232, and 23389+0300.
However, better spectral resolution and S/N are required,
in order to confirm the presence of OF in these objects.

For the remaining object (IRAS 21219-1757), the high values in the FWHM([O III])
are due mainly to the blend of [O III] and Fe II emission lines
(this IR QSO is a strong Fe II emitter).

\end{enumerate}

\vspace{5mm}
\subsection{{\it Wolf Rayet features in a Sample of PG QSOs with Strong
Fe II emission}}

We present in this section the results of the search of Wolf Rayet features
in the large Sample of Bright PG-QSOs (of Boroson \& Green 1992) . This
sample is very interesting since there is a sub-sample of moderate and strong
Fe II emitters, and the authors published the observed spectra with and
without the Fe II contribution (i.e., before and after the subtraction of an
Fe II template). These fact allowed us to perform a careful search of the WR
features in the rest-wavelength range of $\lambda$4600-4700 \AA
(i.e., the strong lines of NII$\lambda$4640 and He II$\lambda$4686).

We found
four Fe II PG-QSOs emitters with  WR features: PG 1244+026, 1448+273,
1444+407 and 1535+547. In Fig. 3 we show the more clear case:
the WR features in PG 1535+547, where the Fe II template was previously
subtracted. In order to study in detail the WR features in these QSOs, we
digitized these four spectra (from Boroson \& Geen 1992).
We discuss these interesting results, in \S 4.2.

\vspace{5mm}
\subsection{{\it [OIII]$\lambda\lambda$4959-5007 in Fe II Emitters (Strong
and Extreme Fe II QSOs)}}

In order to complement the study of multicomponents in the emission line
[O III]$\lambda\lambda$4959-5007 we analyzed the
behavior of this line in our sample of Fe II
extreme (EFE2) and strong (SFE2) emitters (Lipari 1994).

In general, we detected weak [O III]$\lambda\lambda$4959-5007 in strong Fe II systems
(Lipari et al. 1991; Boroson \& Meyer 1992; Lipari et al. 1993, 1994; Lipari
1994). However, we found that only EFE2 of type 1
(i.e. with broad line emission, BLE) show very weak
[O III]$\lambda\lambda$4959-5007 emission (i.e. a ratio
Fe II$\lambda$4925/[O III]$\lambda$4959 $\geq$ 1.0).
The explanation for this fact is complex in the framework of the standard model
for AGNs, since we can see the emission from the BLR but not from the NLR! (the last
is located in the more external parts).

Is important to note, that previously we found  different distributions of SFE2 and
EFE2 emitters (Lipari 1994: Fig. 5), in the IR color diagram
(we define SFE2 and EFE2 as QSOs/galaxies with the emission line ratio of
Fe II$\lambda$4570/H$\beta \geq$ 1 or 2, respectively); which could be
indicative of two QSO populations (see Lipari 1994).
In particular, we found that EFE2 (type 1 and 2) are located
between the power law and black body regions.

For EFE2 systems like PHL 1092, where the width of the BLE is moderate
($\sim$1500 km s$^{-1}$) several works suggested a physical classification
as EFE2 type 2; but this group of type 2 objects shows mainly strong
[O III]$\lambda\lambda$4959-5007 emission, and this is not the case of PHL
1092. Therefore, the [O III] emission could be a good parameter in order to
clarify the properties of each group of EFE2 emitters.
In our starburst scenario, the weak [O III] in IR QSOs is mainly related to
the dust present in these systems. And the presence of ``extreme" Fe II  and
BLE is due -in part- to the presence of multiple SN events.

\vspace{5mm}
\subsection{{\it Morphological Evidence of Outflow, Arcs and Bubbles in Nearby Mergers and IR QSOs}}

In this paper, we report spectroscopic evidence of outflow
in NGC 4039/38, IRAS 01003-2238, 13218+0552, 23128-5919,
11119+3257, 14394+5332, 15130+1958 and 15462-0450.
Previously, we detected  similar outflow features
for the IR mergers NGC 3256, Mrk 231, and IRAS 19254-7245 (Super-Antennae).
In addition, in Mrk 231 we associated the presence of outflow and a
circumnuclear blue arc to a GW scenario; similar to that proposed for
Arp 220 (by Heckman, Armus, \& Miley 1987, 1990).

Recently, high resolution images of IR selected BAL QSOs show in
practically all of these objects the presence of arcs or shells
(Boyce et al. 1997; Stockton, Canalizo, \& Close 1998; Surace et al. 1998; 
Hines et al. 1999) very similar
to those observed in Mrk 231 (the nearest IR merger+GW+Fe {\sc ii}+BAL QSO; 
see Lipari et al. 1994) and in Arp 220 (the nearest IR merger+GW galaxy; see 
Heckman et al. 1987).
These ``circumnuclear and external arcs" could be associated mainly to 
the results of 
interaction of galaxies (tidal tails, rings, etc.) or to the final phase of 
the galactic--wind, i.e., the blowout phase of the galactic bubbles
(Tomisaka \& Ikeuchi 1988; Norman \& Ikeuchi 1989; Suchkov et al. 1994). 
However, for distant AGNs and QSOs it is difficult to discriminate between 
these two related alternatives. 

Even for low redshift BAL IR QSOs -like Mrk 231- there are different 
interpretations about the origin of these ``blue arcs".
In particular, Lipari et al. (1994) found clear evidence of a 
powerful nuclear starburst with galactic--wind in the circumnuclear 
region of Mrk 231, and we proposed a galactic--wind scenario for the origin 
of this blue arc. 
While  Armus, Surace et al. (1994) suggested that this arc has been
originated in the interaction between the main and an obscured second nucleus
(they also suggested that in this blue region and ``shell" there is not
evidence of star-formation process).
Recently, HST--WFPC2(F439W) observations of Mrk 231 
confirm that this blue arc is a {\it ``dense shell of star-forming knots"} 
(see Surace et al. 1998: their Figs. 7 and appendix). 
In addition, these HST broad-band images (Fig. 4b) show blue spiral arms in the
circumnuclear region of Mrk 231 (r $\sim$ 1.5 kpc), similar to those 
observed in the central region of NGC 3256 (Lipari et al. 2000a).
For the remaining selected IR QSOs, showing BAL+Fe {\sc ii} systems, the HST WFPC2 \&
NICMOS high resolution data (Figs. 4) suggest that the observed arcs or shells could be 
related mainly:
(i) in IRAS 04505-2958 and Mrk 231 to star-formation and outflowing material;
(ii) in IRAS 07598+6508 and IRAS 14026+4341 to strong interaction processes; and
(iii) in PG/IRAS 17002+5153 is --for us-- not clear yet 
(see Lipari et al. 2000b).

The galactic-shell in Mrk 231 shows small extension (r $\sim$ 3 kpc).
It is important to study if these giant
galactic-shocks -associated to the compression of the ISM, by the
galactic-shell/wind- could generate new star formation processes.
This mechanisms could produce the
{\it ``dense shell of star-forming knots"} detected in and around the arc
of Mrk 231. And these type of galactic-shocks (associated to the
galactic-wind) show some physical properties similar
to those observed and studied -in detail- in the arms of nearby spirals.

In addition, in  IRAS 19254-7245 (the ``Super-Antennae"),
new HST-WFPC2 broad band images using the filter F814W (Fig. 5a) show a 
complete arc around the southern nucleus, similar to a ``giant SN-ring" with
a r $\sim$4 kpc and an angle to the line of sight of $i_c$ $\sim$50--60$^{\circ}$
(Lipari et al. 2000b). However, this ring was clearly detected only
when the system was located in the PC CCD (when the southern nucleus was
located in the WFC we observed only superposed and confused structures).
In addition, our new NTT high resolution data -obtained for this ULIR merger-
show very extended H$\alpha$ emission in  r $\sim$ 6-7 kpc around the southern
nucleus (i.e., including the region of the ring). In general, this
result is similar to that obtained for NGC 3256 (i.e., extended H$\alpha$
emission in a r $\sim$ 5-6 kpc; see \S 3.1).
It is interesting to note that the ring detected in IRAS 19254-7245 is very
similar to the double arc/shell observed in Arp 220 (Heckman et al. 1987; r
$\sim$ 5 kpc); and the presence of two arcs -in Arp 220- could be explained by
the presence of two compact starburst nuclei.

On the other hand, the presence of ``giant SN-shells or rings"  were
already suggested in order to explain the BAL system in the IR QSO IRAS
07598+6508 (Lipari 1994; cz = 44500 km s$^{-1}$). Heiles (1987) and
Tenorio-Tagle \& Bodenheimer (1988) give references of observational and
theoretical studies of ``giant SN-shells and rings", associated to multiple
explosion of type II SNs.

\section{DISCUSSION}

\subsection{The Galactic-Wind and WR features in NGC 4038/39, IRAS 23128-5919,
and Nearby Mergers}
\vspace{5mm}

Fabbiano et al. (1997), from high resolution X-ray Rosat images, found
extended emission associated to NGC 4039; and they suggested that detailed
spectroscopic observations were required in order to study a
possible nuclear outflow. The result obtained in \S 3.1 shows
the first direct kinematical evidence for nuclear outflow, in NGC 4039.
Which could be only associated to the nuclear starburst in NGC 4039 (and
``galactic--winds"); since, {\it there is not evidence of AGN properties}
in all the multiwavelength studies of The Antennae (including new ISO
observations: Kunze et al. 1997; Fisher et al. 1997; Vigroux et al. 1997).

A similar situation could be explained for IRAS 23128-5919: previous studies
suggested the presence of outflow and WR features, but only the
observations at Casleo, with high S/N, confirmed the presence of
both features in the bright southern nucleus of this IR merger (\S 3.1).

These and previous results for Arp 220, Mrk 266, Mrk 273, NGC 1222, NGC 1614, 
NGC 3256, NGC 3690, NGC 4194, NGC 6240, and other objects (see 
HAM90),  strongly suggest that the relation between {\it merger, 
starburst, galactic--wind, and IR emission}, could play a main role in the evolution 
and formation of galaxies and AGNs (see \S 4.4). 

It is important to note, that mean  values of OF -observed in these and
previous studies of mergers (Lipari et al. 2000a; HAM90)-
are in the velocity range of 100 $\leq$ V$_{OF}$ $\leq$ 700 km s$^{-1}$. And the
situation is different for IR mergers with strong starburts+QSOs (see the next \S).

\vspace{5mm}

\subsection{The Extreme Galactic-Wind and WR features in IRAS 01003-2238,
IRAS 13218+0552 and IR QSOs/galaxies}
\vspace{5mm}

We found EVOF (V$_{OF}$ $\geq$ 1000 km s$^{-1}$)
and/or WR features in 8 objects, including IR mergers, IR QSOs and ULIRGs.
These  objects with EVOF show mainly strong starburst plus the presence
of an obscured IR QSO or AGN; and the nearest objects -Mrk 231 and IRAS 19254-7245-
display clear arcs or shells (associated clearly to ``extreme starburst and GW").
This fact suggests that the interaction between
the two main processes of nuclear activity could generate extreme outflow.
This result is in agreement with a composite model as the source of nuclear
energy in ULIRGs and LIRGs (i.e., starburst plus standard AGN; Perry \& Dyson
1992; Dyson, Perry \& Williams 1992; Perry 1992).

In addition, the presence of strong Wolf Rayet features in several PG QSOs is
indicative of a large number of massive stars (Armus, Heckman \& Miley 1988);
and therefore is indicative -in the future- of the presence of SN of type II.
This fact is also in agreement with a ``composite" model for the source of energy
in these PG QSOs with strong Fe II emission.

\vspace{5mm}
\subsection{The Relation Between: Mergers, Starbursts+ Galactic-Winds, 
and BAL+FeII IR-QSOs}
\vspace{5mm}

The relation between mergers, starbursts and IR QSOs, is important specially for
the study of broad absorption line (BAL) IR selected QSOs. Mainly for the
following reason:
(i) Low et al. (1989), Boroson \& Meyer (1992) found that
IR selected QSOs show a 27$\%$ low-ionization BAL QSO fraction as compared
with 1.4$\%$ for the optically selected high-redshift QSOs sample (Weymann 
et al. 1991);
(ii) {\it extreme} IR galaxies (LIRGs \& ULIRGs) are mainly mergers (\S 1); 
(iii) these objects are also {\it extreme/strong} Fe {\sc ii} emitters 
(Boroson \& Meyer 1992; Lipari 1994);
(iv) Lipari et al. (1994, 1993a); Lawrence et al.
(1997); Egami et al. (1997); and Terlevich, Lipari \& Sodre (2000)
proposed that the {\it extreme} IR+Fe {\sc ii}+BAL phenomena are related --at 
least in part-- to {\it the end phase of an ``extreme starburst" and the 
associated ``powerful galactic--wind"}.
At the end phase of a strong starburst, i.e., type II SN phase
(8-60 $\times$ 10$^{6}$ yr from the initial burst;  
Terlevich et al. 1992; Norman \& Ikeuchi 1989;  Suchkov et al. 1994) 
 naturally appear giant galactic arcs,  
extreme Fe {\sc ii}+BAL systems and dust+IR-emission (Lipari 1994; Lipari et al.
1994, 1993a; Perry \& Dyson 1992; Dyson, Perry, \& Williams 1992; Perry 1992;
Scoville \& Norman 1996; Franco 1999, private communication).

Specifically, in the starburst scenario two main theoretical models for the
origin of BAL systems were proposed:
(i) for IR dusty QSOs/galaxies, in the outflowing gas+dust material the
presence of discrete trails of debris (shed by individual
mass-loss stars) produce the BAL features (Scoville \& Norman 1996); and
(ii) in SN ejecta, which are shock heated when a fast forward shock moves
out into the ISM (with a velocity roughly equal to the ejecta) and a reverse
shock moves and accelerates back towards the explosion center; the
suppression of red-shifted absorption lines arise since SN debris moving
toward the central source are slowed down much more rapidly -by the wind-
than is material moving away (Perry \& Dyson 1992; Perry 1992).
These two alternatives are probably complementary and both explain the main
observed properties of the BAL phenomena (Scoville \& Norman 1996; Perry 1992;
Scoville 1992).
And, in the next paragraphs we will propose a 3rd. scenario for the BAL systems. 

The presence of arcs or shells in IR QSOs could be
a 3rd. (or complementary) explanation -in the starburst
scenario- for the origin of BAL systems in IR QSOs.
The physical processes could be similar to those suggested by Perry (1992)
and Perry \& Dyson (1992), but at larger scale (r $\sim$2-6 kpc).
We found for Mrk 231 -in the blue arc, the nuclear and
circumnuclear regions- a range of expanding velocities ($\Delta$V) from 500 to
8000 km s$^{-1}$ (in the absorption and emission lines; Lipari et al. 1994).
This range is exactly which is required in order to explain the observed BAL
systems and also for theoretical starburst scenarios where 
giant expanding arcs or rings  generate fast giant shocks in the
ISM.
In this model, the high fraction of IR selected QSOs showing
properties of low-ionization
BAL QSO could be explained by the high fraction of arcs, shells, and giant
SN-rings present in these systems (probably originated in the
starburst type II SN phase). 
And in this starburst scenario, the effects of the orientation of the line of
sight and the dust obscuration are also complementary processes
(Perry 1992).

Very recently, detailed new-technology interferometric (IRAM and VLT) and
spectroscopic (ISO) studies with high resolution millimetric, near/mid-IR and 
radio data, confirmed the presence of {\it ``extreme" starburst} in ULIRGs: 
1000 times as many OB stars as 30 Dor in the IR mergers Mrk 231, Arp 
220, Arp 193 (see Downes \& Solomon 1998; Genzel et al. 1998 and Smith et al. 
1998, respectively).
Lipari et al. (2000a), using new-technology data (ESO NTT and HST) combined
with detailed and extensive optical observations (BALEGRE, CASLEO, CTIO) show new 
and clear evidences that NGC 3256 is another example of nearby luminous IR 
merger showing a {\it strong nuclear and extended massive star formation 
process}, with an associated {\it powerful galactic--wind}.

The new detailed results presented in this work (related to extreme GW, \S 3)
give support
to the previous conclusion that {\it the properties of the merger and the 
associated ``extreme" starburst+galactic--wind} play an important role in the 
evolution of LIRGs/ULIRGs and  IR QSOs
(Rieke et al. 1985; Joseph \& Wrigth 1985; Heckman et al. 1987, HAM90;
Lipari et al. 1993a, 1994, 2000a).

In the last years, Thompson, Hill \& Elston (1999)  and
Elston, Thompson, \&  Hill, (1994) reported more than 
15 QSOs at redshift $ 2 <$ z $< 5$, observed at the restwavelength of UV and 
optical Fe {\sc ii}+BAL spectral region. Approximately  50$\%$ of these objects show 
``strong'' Fe {\sc ii} emission, and many of these objects are also BAL+IR QSOs.
In the starburst plus galactic--wind scenario  {\it extreme/strong Fe {\sc ii}+BAL+IR 
emitters are ``young IR QSOs/mergers"} where the starburst is 
probably the dominant source of output energy (Lipari et al. 1994, 1993a; 
Lipari 1994). Therefore, in order to study the real 
nature of these high redshift QSOs it is required a better understanding of the 
merger, starburst, galactic--wind processes in low redshift IR 
galaxies and QSOs (like NGC 3256).

In addition, Thompson et al. (1999) found a lack of iron abundance
evolution in high redshift QSOs: i.e., the absence of increase in the Fe {\sc ii}/Mg {\sc ii}
line ratio and Fe {\sc ii} equivalent width from the earliest epoch (z = 4.47 and
3.35) to the present.
Which represents a problem, since this fact would indicate that 1 Gyr
may be an underestimate of the universe age at z = 4.47
(assuming that SN type Ia is the dominant source of Fe enrichment
in standard models of QSOs); and consequently q$_0$ could be
$\leq$ 0.20 for H$_0$ = 75 km s$^{-1}$ Mpc$^{-1}$. In our proposed starburst
scenario, both results: the detection of strong Fe {\sc ii} in QSOs at redshift
2 $\leq$ z $\leq$ 5 (or even at z $\geq$ 5),
and the lack of iron abundance evolution in high redshift QSOs agrees
with the prediction of our models (Lipari et al. 1993a, 1994;
Lipari 1994; Terlevich et al. 1992, 2000). Specifically, since 
in our scenario the time for the strong Fe enrichment in the shell
of SN type II and in the ISM is $\sim$ 8-60 $\times$10$^6$
yr (i.e., the end phase of an ``extreme starburst"). This time
is very short in relation to that required for the Fe enrichment
of the ISM by SN type Ia: i.e., 2 $\times$10$^9$ yr
(see Friaca \& Terlevich 1998). Therefore --in this scenario-- the results
obtained by Thompson et al. (1999)  do not represent a problem
with the present accepted age of the universe at z $\sim$ 5:
 $\sim$10$^{9}$yr for q$_0$ = 0.5 and H$_0$ = 75 km s$^{-1}$
Mpc$^{-1}$.

The {\it evolutive} end product of this interaction between 
mergers and extreme starburst+GW  could be: 
(i) SMBHs and IR-QSOs in the nuclear region, according to 
the conditions of the merger/starburst processes, such as the
nuclear compression of the ISM gas, the inflow and outflow rate, etc
(Genzel et al. 1998; Downes \& Solomon 1998; Taniguchi et al. 1999;
Lipari et al. 2000a);
and (ii) elliptical, cD, radio galaxies for the multiple-merger 
process as a whole (Toomre 1977; Schweizer 1978, 1982; Sanders et al. 1988a; 
Barnes 1989; Barnes \& Hernquist 1992; Shier, Rieke, \& Rieke 1994, 1996;
Weil \& Hernquist 1996).

\vspace{5mm}
\subsection{The Relation between Mergers+ Starbursts/ galactic-winds and
 the Formation and Evolution of Galaxies}
\vspace{5mm}

The results obtained in \S 3 are also interesting in the study 
of high redshift objects and {\it galaxies formation,} since
it is expected that the properties of the {\it  initial collapse, merger}
and {\it starburst+galactic--wind} play an important role, in practically
all the scenarios of galaxy formation  (see Larson 1974; Ostriker \&
Cowie 1981; Dekel \& Silk 1986; Ikeuchy \& Ostriker 1986; Lacey \& Silk 1991;
 Berman \& Suchkov 1991; Cole et al. 1994).
In particular, in luminous IR mergers the SFRs are close to 
those inferred of a galaxy forming itself: IR luminosities of $\sim$ 
10$^{11-12}$ L$_{\odot}$ imply SFRs of $\sim$300--500 M$_{\odot}$ yr$^{-1}$,
if such SFRs are sustained for galaxy free-fall times  10$^{8-9}$ yr$^{-1}$, 
the total mass of newly formed stars would be 10$^{10-11}$  M$_{\odot}$ 
(see HAM90). 

When we observe locally (in NGC 3256, Arp 220, Mrk 231, and
others) the galactic--wind in luminous IR mergers,
 we are probably observing the feedback processes from
massive star formation that may have important influence in determining the 
overall structure of galaxies in the {\it ``general dissipative collapse"} 
(Rees \& Ostriker 1977; Silk 1977; Martin 1999; 
Norman \& Ikeuchy 1989; Kormendy \& Sanders 1992; Bekki \& Shioya 1998;
HAM90; Tenorio-Tagle, Rozyczka, \& Bodenheimer 1990;  
Chevalier \& Clegg 1985).
More specifically, 
in the early stage of galaxy formation (when the SFR is expected to be
higher) the {\it ``galactic-wind"} plays a decisive role in the 
feedback process: reheating the ISM, and contributing to stop the 
initial collapse, and therefore would determine the overall structure of galaxies. 
The {\it ``galactic-wind"} also plays a central role in  
some particular galaxy formation scenarios, for example in the
{\it ``explosive" and ``hot" scenarios} (postulated by Ostriker \& Cowie
1981 and Berman \& Suchkov 1991, respectively) where the SN explosions and
galactic--wind are the process of SFR self-regulation  in young galaxies
(see also McKee \& Ostriker 1977; HAM90; Lipari et al. 1994).
And, these properties are very similar to those proposed for Mrk 231 and
``The Super-Antennae" (see \S 3, and Lipari et al. 2000a), in the observed
extended massive star formation and the {\it ``extreme" galactic--wind
processes (with their associated shells/arcs)}.

SNs of type II  are highly concentrated in space and time, and arise from
massive stars (m$\geq$ 5 M$_{\odot}$) in young stellar clusters and
associations  (of tens or hundreds; Heiles 1987).
And giant galactic-shells have been detected (see for references Heiles
1992; and Tenorio-Tagle \& Bodenheimer 1988).
In addition, the presence of {\it ``extreme" starburst and
the associated SN events} in ULIRGs (Downes \& Solomon 1998; Genzel et al. 1998
and Smith et al. 1998) is also a confirmation of the existence of ``multiple"
type II SN explosions". These
 ``multiple" type II SN explosions are the main galactic process capable to generate the
blow-out phase of the galactic--winds (arcs and shells),
large amounts of dust and IR-emission,
Fe overabundance, and the BAL phenomena (Norman \& Ikeuchi 1989; Perry \&
Dyson 1992; Lipari et al. 1993a, 1994; Lipari 1994; Scoville \& Norman 1996).
However,
in the dusty nuclear regions of LIRGs and ULIRGs (with A$_{V}$ $\sim$10-1000 
mag; see Sakamoto et al. 1999; Genzel et al. 1998), the presence of
type II super/hyper--nova could be detected only for nearby systems
and using interferometric radio data (Smith et al. 1998).

New submillimeter-wavelength surveys show a population of
very dusty star forming galaxies at high redshifth (Hughes et al. 1998;
Berger et al. 1998), with very similar properties to those observed in
LIRGs and ULIRGs (Scott 1998).
Therefore, in order to study distant IR mergers, objects
with composite source of nuclear energy (QSOs plus starburst), and very
dusty galaxies, the more clear signals of ``extreme starbursts" are 
the above described features associated to the presence of powerful galactic--winds
and ``multiple" type II SN explosions
(e.g., galactic-shells/arcs, spectra with outflow or WR components,
very blue spiral arms, ``extreme" amount of dust and abundance of Fe II, etc).
Those features are similar to those observed --at low redshift-- in Arp 220, Mrk 231,
NGC 3256, IRAS 19254--7245, IRAS 07598+6508, IRAS 22419--6049, IRAS 04505--2958
and others.

\acknowledgments

The authors thank P. Amram, T. Boroson, J. Boulesteix, R. Green,
L. Hernquist, and J. Hibbard for discussions and help.
We would like to express our gratitude to the staff members and observing 
assistants at BALEGRE, Casleo, CTIO and ESO Observatories. 
This work was based on observations made using the NASA and ESA HST and IUE
satellite, obtained from archive data at ESO-Garching and 
STScI-Baltimore. This work was made using the NASA Extragalactic Database NED;
which  is operated by the Jet Propulsion Laboratory, California Inst. of 
Technology, under contract with NASA. 
This work was supported in part by Grants from Conicet, Conicor, 
Secyt-UNC, and Fundaci\'on Antorchas (Argentina).

\newpage

\centerline{\bf REFERENCES}

\vspace{5mm}

\noindent
Afanasiev, V., Dodonov, S., \& Carranza, G. 1994, Bol. Asoc. Argentina de 

Astron. 39, 160\\
Armus, L., Heckman, T.M.,  \& Miley, G. 1988, ApJ, 326, L45 \\
Barnes, J., \& Hernquist, L. 1992, ARA\&A, 30, 705 \\
Beichman, C. A. et al. 1986, ApJ, 308, L1 \\
Bekki, K. \& Shioya, Y. 1998, ApJ, 497, 108 \\
Bergvall, N. \& Johansson, L.  1988, A\&A, 149, 475 \\
Berger, A. J.  et al. 1998, Nature, 394, 248 \\
Berman, B., \& Suchkov, A. 1991, Astrophy. \& Space Sc., 184, 169\\
Bevington, P. 1969, Data Reduction and Error Analysis for the Physical

 Sciences, (New York: McGraw-Hill)\\
Boroson, T., \& Green, R.  1992, ApJS, 80, 109 \\
Boroson, T., \& Meyer, K.  1992, ApJ, 397, 442\\
Boyce, P. J., et al.  1996, ApJ, 473, 760\\
Canalizo, G., Stockton, A., \& Roth, K.  1998, AJ, 115, 890 \\
Carral, P., et al.  1994, ApJ, 423, 223 \\
Chevalier, R., \&  Clegg, A. 1985, Nature, 317, 44 \\
Clements, D. et al. 1996, MNRAS, 279, 459 \\
Cole, S., et al. 1994, MNRAS, 271, 181 \\
Colina, L., Lipari, S. L., \& Macchetto, F.  1991a, ApJ, 379, 113 \\
Colina, L., Lipari, S. L., \& Macchetto, F.  1991b, ApJ, 382, L63 \\
Condon, J. Anderson, M., \& Helow G. 1991a, ApJ, 376, 95 \\
Condon, J., Huang, Z., Yin, Q. \& Thuan, T.  1991b, ApJ, 378, 65 \\
Dekel, A. \& Silk, J. 1986, ApJ, 303, 39 \\ 
Diaz, R., Carranza, G., Dottori, H. \& Goldes, G. 1999, ApJ, 512, 623\\ 
Djorgovski, S. 1994, in Proc. Conf. Mass-Transfer Induced Activity in 

Galaxies, ed. I. Shlosman (Cambridge: Cambridge University Press), 452\\
Downes, D., \& Solomon, P. M. 1998, ApJ, 507, 615\\
Dyson, J., Perry, J., \& Williams, R. 1992, in Testing the AGN Paradigm,

ed. S. Holt, S. Neff, \& M. Urry (New York: AIP), 548\\
Elston, R., Thompson, K., \& Hill, J. 1994, Nature, 367, 250 \\
Fabbiano, G., Schweizer, F., \& Mackie, G. 1997, ApJ, 478, 542 \\
Franco, J. \& Ferrara, A. 1992, in Proc. Evolution of ISM and Dynam. of Galax.

eds. J.Palous, W. Burton, P. Lindblad (Cambridge: Camb. Univ. Press), 130\\
Friaca, A., \& Terlevich, R. 1998, MNRAS, 298, 399\\
Fischer, J. et al. 1996, A\&A, 315, L97\\
Genzel, R., et al. 1998, ApJ, 498, 579\\
Hibbar, J. E.,  et al. 1994, AJ, 107, 67\\
Hines, D.,  et al. 1999, ApJ, 512, 140\\
Heckman, T.M., Armus, L., \& Miley, G. 1987, AJ, 93, 276\\
Heckman, T.M., Armus, L., \& Miley, G. 1990, ApJS, 74, 833 (HAM90)\\
Heiles, C. 1987, ApJ,315, 555\\
Heiles, C. 1992, in Proc. Evolution of ISM and Dynamics of Galaxies, eds.

Hughes, D. H. et al. 1998, Nature, 394, 241 \\
Hutching, J. B.,  \& Neff, S. 1991, AJ, 101, 434\\
Ikeuchi, S. \&  Ostriker, J. 1988, ApJ, 301, 522 \\
Johansson, L. \& Bergvall, N. 1988, A\&A, 192, 81 \\
Joseph, R. D.,  \& Wright, G. 1985, MNRAS, 214, 87\\
Kim, D.,  \& Sanders, D. 1998, ApJS, 119, 41\\
Kim, D., Veilleux, S., \& Sanders, D. 1998, ApJ, 508, 627\\
Kormendy, J., \& Sanders, D. 1992, ApJ, 390, L53\\
Kunze, D. et al. 1996, A\&A, 315, L101\\
Lacey, C. \& Silk, J. 1991, ApJ, 381, 14 \\
Larson, R.  1974, MNRAS, 166, 585 \\
Lawrence, A., et al.  1988, MNRAS, 235, 261 \\
Lawrence, A., et al.  1997, MNRAS, 285, 879 \\
Leitherer, C. 1991, in Massive Stars in Starbursts, eds. C. Leitherer, 

N. Walborn, T.M. Heckman, C.A. Norman (Cambridge Univ. Press), 1\\
Leitherer, C., \& Heckman, T. 1995, ApJS, 96, 9 \\
Lipari, S. L. 1994, ApJ, 436, 102 \\
Lipari, S. L., Colina, L., \& Macchetto, F.  1994, ApJ, 427, 174 \\
Lipari, S. L., et al. 2000a, AJ, in press (astro-ph 9911019) \\
Lipari, S. L., et al. 2000b, in preparation \\
Lipari, S. L., \& Macchetto, F.  1992, ApJ, 387, 522 \\
Lipari, S. L., Macchetto, F., \& Golombeck, D. 1991, ApJ, 366, L65 \\
Lipari, S. L., Terlevich, R., \& Macchetto, F.  1993a, ApJ, 406, 451 \\
Lipari, S. L., Tsvetanov, Z., \& Macchetto, F.  1993b, ApJ, 405, 586 \\
Lipari, S. L., Tsvetanov, Z., \& Macchetto, F.  1997, ApJS, 111, 369 \\
Low, F., Cutri, R., Huchra, J., \& Kleinmann, S., 1988, ApJ, 327, L41 \\
Low, F., Cutri, R., Kleinmann, S., \& Huchra, J.  1989, ApJ, 340, L1 \\
Lutz, D., Veilleux, S., \& Genzel, R. ApJ, 517, L13\\
Lutz, D., et al. 1996, A\&A, 315, L137\\
Marochnik, L., \& Suchkov, A. 1992, in Proc. Evolution of ISM and Dynam. of

Galax., eds. J.Palous, W.Burton, P.Lindbland(Cambridge: Camb.Univ.Press), 234\\
Martin, C. 1999, ApJ, 513, 156 \\
Mathews, W. G., \& Dones J. 1992, Lick Observatory Bull., preprint \\
Mc-Carthy, P., Heckman, T., \& van Breugel, W. 1987, AJ, 93, 264\\
McKee, C.,  \& Ostriker, J. 1977, ApJ, 218, 148 \\
Melnick, J., \& Mirabel, I. F. 1990, A\&A, 231, L19\\
Mihos, C., \& Hernquist, L. 1994a, ApJ, 425, L13\\
Mihos, C., \& Hernquist, L. 1994b, ApJ, 431, L9\\
Mihos, C., \& Hernquist, L. 1996, ApJ, 464, 641\\
Mirabel, I. F., Dottori, H., \& Lutz, D. 1992, A\&A, 256, L19\\
Mirabel, I. F., Lutz, D., \& Maza, J. 1991, A\&A, 243, 367\\
Noguchi, M. 1991, MNRAS, 251, 360 \\
Norman, C., \& Ikeuchi, S. 1989, ApJ, 395, 372\\
Norman, C., \& Scoville, N. 1988, ApJ, 332, 124\\
Okumura S., et al. 1991, in IAU Symposium 146, Dynamics of Galaxies and 

their Molecular Distributions, ed. F. Combes (Dordretch: Kluwer), 425\\
Osterbrock, D. 1989, in Astrophysics of Gaseous Nebulae and Active 

Galactic Nuclei, (University Science Book: Mill Valley)\\
Ostriker, J., \& Cowie, L. 1981, ApJ, 243, L127 \\
Perry, J. 1992, in Relationships Between AGN and Starburst

Galaxies, ed. A. Filippenko (San Francisco: ASP Conf.S.31), 169\\
Perry, J., \& Dyson, R. 1992, in Testing the AGN Paradigm, ed.

S. Holt, S. Neff, \& M. Urry (New York: AIP), 553\\
Read, A., Ponman, T., \& Wolstencroft, R.1995, MNRAS, 277, 397 \\
Remillard, et al. 1993, AJ, 105, 2079 \\
Rees, M., 1977, ARA\&A, 42, 471 \\
Rees, M.,  \& Ostriker, J. 1977, MNRAS, 179, 541 \\
Rieke, G.,  et al. 1985, ApJ, 290, 116 \\
Rieke, G.,  et al. 1980, ApJ, 238, 24\\
Rigopoulou, D., et al. 1996, A\&A, 315, L125 \\
Rosa, M., \& D'Oddorico, S. 1986, in IAU Symposium 116, ed. C. de Loore, 

A. Willis \& P. Laskarides (D. Reidel P. C.), 355 \\
Sanders, D. B., Egami, E., Lipari, S., Mirabel, I., \& Soifer B. T. 1995, 

AJ, 110, 1993 \\
Sanders, D. B., \& Mirabel, F. 1996, ARA\&A, 34, 749\\
Sanders, D. B., Scoville, N., \& Soifer B. T. 1991, ApJ, 370, 158 \\
Sanders, D.B., Soifer, B.T., Elias, J.H., Madore, B.F., Matthews, K.,

 Neugebauer, G., \& Scoville, N.Z. 1988a, ApJ, 325, 74\\
Sanders, D.B., Soifer, B.T., Elias, J.H., Neugebauer, G., \& Matthews, K.

 1988b, ApJ, 328, L35\\
Scott, D.,  et al. 1998, Nature, 394, 219\\
Scoville, N. Z. 1992, in Relationships Between AGN and Starburst

Galaxies, ed. A. Filippenko (San Francisco: ASP Conf.S.31), 159\\
Scoville, N. Z., \& Norman, C. 1996, ApJ, 451, 510 \\
Scoville, N. Z., Sargent, A., Sanders, D., \& Soifer, B. 1991, ApJ, 366, L5 \\ 
Scoville, N. Z., \& Soifer, B.T. 1991, in Massive Stars in Starbursts, eds. C. 

Leitherer, N. Walborn, T.M. Heckman, C. Norman (Cambridge Univ. Press), 233\\  
Silk, J. 1977, ApJ, 211, 638\\
Schweizer, F.  1980, ApJ, 237, 303\\
Schweizer, F.  1982, ApJ, 252, 455\\
Schweizer, F. et al. 1996, AJ, 111, 109 \\
Smith, H., Lonsdale, C., Lonsdale, C., \& Diamond, P. 1998, ApJ, 493, L17\\
Soifer, B.T., et al. 1984, ApJ, 283, L1\\
Soifer, B.T., et al. 1987, ApJ, 320, 238\\
Stockton, A., Canalizo, G., \& Close, L. 1998, ApJ, 500, L121\\
Stone, R.,  \& Baldwin, J. 1983, MNRAS, 204, 357\\
Suchkov, A., Allen, R., \& Heckman, T. 1993, ApJ, 413, 542\\   
Suchkov, A., Balsara, D., Heckman, T., \& Leitherer, C. 1994, ApJ, 430, 511\\   
Suchkov, A., \& Berman, V. 1992, in Proc.Evolution of ISM and Dynam. of Galax.

eds. J.Palous, W. Burton, and P. Lindblad (Cambridge: Camb. Univ. Press), 392\\
Suchkov, A., Berman, V., Heckman, T., \& Balsara, D 1996, ApJ, 463, 528\\   
Surace, J., et al. 1998, ApJ, 492, 116\\
Taniguchi, Y., Ikeuchi, S., \& Shioya, K. 1999, ApJ, 514, L9\\
Taniguchi, Y., \& Shioya, K. 1998, ApJ, 501, L67\\
Taniguchi, Y., Trentham, N., \& Shioya, K. 1998, ApJ, 504, L79\\
Taniguchi, Y., \& Wada, K. 1996, ApJ, 469, 581\\
Tenorio-Tagle, G., \& Bodenheimer, P. 1988, ARA\&A, 26, 145\\
Terlevich, R., \& Boyle, B. 1993, MNRAS, 262, 491\\
Terlevich, R., et al. 1992, MNRAS, 255, 713\\
Terlevich, R., Lipari, S., \& Sodre, L. 2000, MNRAS, in preparation\\
Thompson, K., Hill, J., \& Elston,  R. 1999, ApJ, 515, 487 \\  
Tomisaka, K., \& Ikeuchi, S. 1988, ApJ, 330, 695 \\  
Vader, J. P. et al. 1987, AJ, 94, 847 \\
Veilleux, S., et al. 1996, ApJS, 98, 171\\
Veilleux, S., Kim, D., \& Sanders, D. 1999, ApJ, 522, 113\\
Vigroux, L. et al. 1996, A\&A, 315, L93\\

\newpage

\centerline{\bf FIGURE CAPTIONS}

\vspace{7mm}

\noindent
{\bf Fig. 1.}
{\bf (a)} 
CASLEO optical spectrum of NGC 4039 in the H$\alpha$+[N {\sc ii}]${\lambda\lambda6548,6583}$
region and along PA 174$^{\circ}$, through the nucleus. It shows 
 the OF nuclear component. 

{\bf (b-d)} 
MKO optical spectra of the QSOs IRAS 01003-2238, IRAS 13218+0552 and
CASLEO sectra of IRAS 19254-7245. In the H$\beta$+[O {\sc ii}]
${\lambda\lambda4959,5007}$ region, showing the EVOF components. 

\vspace{3mm}

\noindent
{\bf Fig. 2. }
MKO optical spectrum of the  IRAS 11119+3257 in the H$\beta$+[O {\sc ii}]
${\lambda\lambda4959,5007}$ region, showing the EVOF components. 

\vspace{3mm}

\noindent
{\bf Fig. 3.}
Optical spectrum of the QSOs  PG 1535+547, showing WR features at
NII$\lambda$4640 and He II$\lambda$4686. The spectrum was digitized
from Boroson \& Green (1992), and a Fe II template was subtracted.

\vspace{3mm}

\noindent
{\bf Fig. 4.}
HST broad--band images of 8 strong IR+Fe II QSOs
(the first 4 objects are also BAL QSOs),
the  N-S lines are rotated, from the top.
Note that all these IR objects show ``arcs or shells" and/or merger features.

{\bf (a-d)} 
IRAS 07598+6508, IRAS 12540+5708 (Mrk 231), IRAS 14026+4341, IRAS/PG 1700+518.

{\bf (e-h)} 
IRAS 04505-2958, IRAS 00275-2859, IRAS 13349+2438 and I ZW 1.

\vspace{3mm}

\noindent
{\bf Fig. 5.}
HST broad--band images of luminous IR QSOs with OF
(the  N-S lines are rotated, from the top).

{\bf (a-c)} 
IRAS 19254-7245, IRAS 13218+0552 and  IRAS 23128-5919.

$${
\begin{tabular}{lcccc}
\multicolumn{5}{c}{\bf Table 1: Extreme velocity outflow in IR QSOs}\\
\\
\hline
\hline

Object (IRAS)        & V$_{OF1}$   & V$_{OF2}$   & cz          & Comments \\
                     & km s$^{-1}$ & km s$^{-1}$ & km s$^{-1}$ &    \\
\hline
                     &             &             &             &    \\

01003-2230           & -1530       &  -710       &  35505      &  at O III  \\
13218+0552           & -1850       &  ---        &  61000      &   " \\
19254-7245           & -1000       &  ---        &  17900      &   " \\
Mrk 231              & -1000       &  ---        &  12670      &  at O II  \\
                     &             &             &             &    \\
11119+3257           & -2120       &  -1330      &  56230      &  at O III  \\
14394+5332           & -1750       &   -850      &  31475      &    "\\
15130+1958           & -1380       &   -890      &  32700      &    "\\
15462+0450           & -1800       &   -970      &  30030      &    "\\
                     &             &             &             &    \\
\hline

\end{tabular}
}$$

\noindent
                                   
\end{document}